\documentclass[preprint,1p]{elsarticle}

\usepackage{lmodern}
\usepackage[T1]{fontenc}
\usepackage{amsmath,amsfonts,amssymb}
\usepackage{graphicx}
\usepackage{booktabs}
\usepackage{multirow}
\biboptions{numbers,sort&compress}
\usepackage{hyperref}
\hypersetup{bookmarksnumbered,%
    colorlinks,%
    linkcolor=blue,%
    citecolor=blue,%
    plainpages=false,%
    pdfstartview=FitH}

\journal{Nuclear Physics A}

\begin{document}

\begin{frontmatter}

\title{$\alpha$-decay widths from two-potential approaches and WKB prefactors}

\author[1,2]{Nguyen Thi Huyen Nga}

\author[1,2]{Le Quoc Thai}

\author[1,2]{Chau Van Tao}

\author[1,2]{Le Hoang Chien}

\author[1,2]{Nguyen Tri Toan Phuc\corref{cor1}}
\ead{nttphuc@hcmus.edu.vn}

\cortext[cor1]{Corresponding author}

\affiliation[1]{organization={Department of Nuclear Physics, Faculty of Physics and Engineering Physics, University of Science},
            city={Ho Chi Minh City},
            country={Vietnam}}

\affiliation[2]{organization={Vietnam National University},
            city={Ho Chi Minh City},
            country={Vietnam}}

\begin{abstract}
The accurate calculation of $\alpha$-decay widths can be obtained with several equivalent descriptions of the Gurvitz-Kalbermann two-potential approach (TPA) or a semiclassical WKB description having both correct barrier penetrability and consistent inner-state normalization prefactor. This prefactor, which is related to the assault frequency, is implemented in several ways in the literature, producing systematically different results that can be mistaken for $\alpha$ preformation effects. We benchmark single-particle $\alpha$-decay widths across 161 even-even nuclei ($A_p=144$--$252$) using several TPA descriptions and WKB calculations with four common prefactors, employing a Woods-Saxon potential fixed by the quantization-constrained parameter method. The TPA formulas agree to within a few parts in $10^{-8}$, highlighting their robustness. Among semiclassical prefactors, only the widely used round-trip (RT) approximation closely reproduces the TPA widths to within a few percent. An Airy-function analysis reveals this $\cos^2\!\to\!1/2$ averaged WKB prefactor is the correct leading-order semiclassical normalization of the exact quasibound state, clarifying the counterintuitive empirical observations associated with the WKB prefactors. Our benchmark identifies the WKB prefactor appropriate for $\alpha$-decay tunneling analysis and provides a clean reference for isolating preformation and structure effects in future studies.
\end{abstract}



\end{frontmatter}

\section{Introduction}\label{sec:intro}

Alpha decay is a fundamental quantum tunneling process in nuclear physics and plays a central role in the study of heavy and superheavy nuclei \cite{Gamow1928,Gurney1929,Smits2024SHE,Qi2019,Belli2019,Delion2023,Wei2024}. Measured lifetimes and branching ratios constrain cluster formation, shell effects, and nuclear structure at the limits of stability. Microscopic quartetting and pocket-potential studies confirm that $\alpha$ clustering is predominantly a surface phenomenon governed by shell structure, Pauli blocking, and medium effects \cite{Ropke2014,Xu2016,Yang2021Shell,Zheng2024Pocket}. This connection is particularly important for current superheavy-element (SHE) programs at JINR, RIKEN, GSI, and related facilities, where new isotopes are routinely established through correlated $\alpha$-decay chains \cite{Smits2024SHE,Roberto2023ActinideTargets,Belli2019}. The experimentally measured half-life inherently includes contributions from both the $\alpha$ preformation within the parent nucleus and its subsequent tunneling probability through the Coulomb barrier. Because the preformation factor is strongly model dependent \cite{Qi2019,Delion2023,Lovas1998}, deviations from experiment can be attributed to preformation or structure effects only after the tunneling width itself has been computed under controlled theoretical assumptions and approximations. Establishing such a controlled theoretical baseline is the primary purpose of this work.

The semiclassical Wentzel-Kramers-Brillouin (WKB) approximation has been a primary tool for systematic $\alpha$-decay studies for decades due to its broad applicability and low computational cost \cite{Buck1992,Mohr2006,Pei2007,Ismail2010,Kelkar2007,CHIEN2022}. However, while the action integral robustly determines the exponential penetrability, the absolute width also requires an assault frequency (or a normalization prefactor) that encodes the motion of the quasibound state in the inner pocket. On this point, there is a lack of consensus among the prefactor descriptions used in $\alpha$-decay studies. Kelkar and Casta\~neda emphasized that WKB decay widths are meaningful only when the quantization condition, the penetrability, and the prefactor are implemented consistently within a single semiclassical framework \cite{Kelkar2007}. In practice, round-trip-period, harmonic-oscillator, and classical-frequency prescriptions all remain in active use \cite{CHIEN2022,Sharma2022frequency,Bao2014GLDM,Deng2021GLDM,Kelkar2007}, while the semiclassical reduction of the two-potential approach by Gurvitz and Kalbermann supplies a more foundational normalization, built from the form factor of the WKB wave in the pocket \cite{Gurvitz1987}. Because all of these descriptions share the same penetrability, the choice of prefactor uniformly shifts every width in a calculation by a common factor. The practical question is therefore not whether WKB is an accurate tool for systematic $\alpha$-decay analysis, but which prefactor correctly reproduces the absolute width of a fully quantum calculation and why it does so.

Fully quantum-mechanical approaches for tunneling decay resolve the prefactor ambiguity by construction, because the normalization of the decaying state is consistently determined from the wave function rather than prescribed. The two-potential approach (TPA) of Gurvitz and Kalbermann separates the interaction into an inner quasibound problem and an outer scattering problem, yielding the decay width in a compact analytical form \cite{Gurvitz1987,Gurvitz1988}. The four Gurvitz-Kalbermann (GK) formulas used below are the volume, surface, and boundary representations of this single construction. Gurvitz and collaborators later proposed the modified two-potential approach (MTPA), which replaces the sharp potential split with a controlled matching scheme in the barrier region \cite{Gurvitz2004}. The reliability of these quantum formulations has been tested in several important comparisons. \AA{}berg \textit{et al.} demonstrated that TPA widths are practically identical to distorted-wave (DW) method \cite{Kadmensky1970} widths for spherical proton emitters, and their erratum recorded the perplexing empirical fact that the simple averaged (round-trip) WKB normalization, rather than the more elaborate one from GK matching, gives the closest agreement with DW results \cite{Aberg1997,Aberg1998Erratum}. More recently, Alavi \textit{et al.} \cite{Alavi2024} carried out the analogous comparison for uranium $\alpha$ decay with a microscopic double-folding interaction, finding very good agreement among TPA, direct, and DW widths and showing that the TPA result is insensitive to the choice of separation radius. More recently, Bo shows that the WKB can reproduce the $\alpha$-decay half-lives data with comparable accuracy with the DW method while being less sensitive to the potential parameters \cite{Bo2026}. The present work extends this line of investigation by broadening the quantum TPA methods comparison from a single isotopic chain to the even-even $\alpha$-decay landscape and by confronting the resulting TPA reference with the commonly used semiclassical prefactor families in the WKB method.

A spherical baseline of this kind is also timely. Recent studies have pushed the tunneling problem well beyond the spherical TPA picture. Li and Xu formulated a three-dimensional TPA (3D-TPA) for deformed $\alpha$ emitters \cite{Li2023Alpha3D}, Zheng \textit{et al.} connected pocket formation and $\alpha$ clustering to nuclear-medium effects and applied a deformed pocket-type dynamical double-folding potential with deformed TPA widths \cite{Zheng2024Pocket,Zheng2025Deformation}, You \textit{et al.} surveyed even-even nuclei with a deformed TPA \cite{You2025DeformedTPA}, while Cheng and Xu combined 3D-TPA with physics-informed neural networks \cite{Cheng2026}. Before deformation, channel coupling, or microscopic preformation factors are layered onto the calculation, however, the method dependence of the single-particle width for a fixed spherical input should be pinned down, since otherwise effects at the level of tens of percent can be mistaken for structure properties.

This work accordingly pursues three aims. The first is to verify that the four Gurvitz TPA formulas and the MTPA matching expression define the same single-particle $\alpha$ width for a fixed Woods-Saxon (WS) input, thereby establishing a stable quantum reference over a broad even-even dataset. The second is to identify which WKB prefactor preserves that quantum width closely enough for practical surveys. The third is to explain the observed discrepancies between these descriptions. An Airy-function analysis of the quasibound-state norm shows that the familiar $\cos^2\!\to\!1/2$ averaging is not a crude shortcut but the correct leading-order semiclassical normalization of the exact state, whereas the literal Gurvitz-Kalbermann prescription retains a spurious turning-point contribution evaluated precisely where the WKB amplitude breaks down. The long-noticed and surprising empirical agreement of the averaged round-trip prefactor with the TPA in Ref.~\cite{Aberg1998Erratum} is revisited and explained in light of Airy-function analysis.

All calculations use a WS interaction fixed by the quantization-constrained parameter method (QCPM) introduced in Ref.~\cite{Nga2026}. The QCPM-derived WS potential serves as a fixed common input throughout, so that the comparison isolates the part of the single-particle width scale generated by the tunneling, matching, normalization, and radial quantization conventions themselves rather than by any width or preformation fitting. The global comparison covers 161 even-even emitters with parent mass numbers $A_p=144$--252, supplemented by a separate 10-nucleus run spanning $A_p=104$--294 as an illustrative benchmark over a wider region of the nuclear chart. Section~\ref{sec:theory} presents the formalism, Sec.~\ref{sec:results} the numerical comparison and the Airy analysis of the prefactor hierarchy, and Sec.~\ref{sec:conclusion} the conclusions.

\section{Formalism}\label{sec:theory}

\subsection{Interaction potential}\label{subsec:potential}
We consider the $\alpha$-daughter relative motion in a central potential, and this same spherical input defines the benchmark used throughout the paper. For the TPA and MTPA calculations the one-body interaction takes the form
\begin{equation}\label{eq:vtotal}
V(r)=V_N(r)+V_C(r)+\frac{\hbar^2\ell(\ell+1)}{2\mu r^2},
\end{equation}
where $\mu$ is the reduced mass of the $\alpha$-daughter system and $\ell$ is the orbital angular momentum quantum number. In the present work, all decay processes are even-even favored ground-state transitions evaluated with $\ell=0$. The Coulomb interaction takes the standard point-sphere form
\begin{equation}\label{eq:coulomb}
V_C(r)=
\begin{cases}
Z\left(3-r^2/R_0^2\right)/(2R_0), & r<R_0,\\[4pt]
Z/r, & r\ge R_0.
\end{cases}
\end{equation}
Here $Z=Z_cZ_de^2$, with $Z_c=2$ is the $\alpha$-particle charge number and $Z_d$ is the daughter charge number. The nuclear interaction takes the standard WS form
\begin{equation}\label{eq:ws}
V_N(r)=\frac{-V_0}{1+\exp[(r-R_0)/a]},
\end{equation}
with depth $V_0$, radius parameter $R_0$, and diffuseness $a$.

Rather than adjusting $V_0$ and $a$ directly to decay observables, we reduce the ambiguity in choosing the potential shape by using the spherical form of the quantization-constrained parameter method (QCPM) \cite{Nga2026}. The WS radius equals the parent nuclear radius, $R_0=R_p$, with
\begin{align}\label{eq:radii}
R_p = 1.226A_p^{1/3}+2.86A_p^{-2/3}+0.99\frac{\Delta E_p}{A_p} -1.09\frac{(A_p-2Z_p)2Z_p}{A_p^2},
\end{align}
where the radius formula is taken from Refs.~\cite{Wang2013,Wang2010}, $A_p$ and $Z_p$ are the parent mass and charge numbers, and $\Delta E_p$ is the shell-correction energy.

The QCPM quantization condition uses the Langer centrifugal coefficient \cite{Langer1937,Berry1972}
\begin{equation}\label{eq:clanger}
C_\ell=\frac{\hbar^2}{2\mu}\left(\ell+\frac{1}{2}\right)^2,
\end{equation}
as required for radial semiclassical quantization. The same convention enters the WKB calculation below, whereas the TPA/MTPA widths retain the physical centrifugal term $\hbar^2\ell(\ell+1)/(2\mu r^2)$ in Eq.~(\ref{eq:vtotal}). By matching the potential derivative at the barrier top, the depth is related to the diffuseness and the barrier-top radius $R_b$ through \cite{Nga2026}
\begin{equation}\label{eq:v0}
V_0=\frac{Za}{R_b^2}\,
\frac{\left[1+\exp[(R_b-R_0)/a]\right]^2}{\exp[(R_b-R_0)/a]}.
\end{equation}
The barrier-top position follows from the barrier height $V_b$,
\begin{equation}\label{eq:rb}
R_b=\frac{1}{2}\left[\frac{Z}{V_b}+\sqrt{\left(\frac{Z}{V_b}\right)^2
-4\frac{(Za-C_\ell)}{V_b}}\right].
\end{equation}
The barrier height $V_b$ is taken from the global $\alpha$-nucleus potential of Avrigeanu \textit{et al.} evaluated at the experimental decay energy \cite{Avrigeanu2014,Nga2026}. From Eq.~(\ref{eq:rb}), the diffuseness $a$ can be determined, followed by the depth $V_0$ from  Eq.~(\ref{eq:v0}). 

For the WKB calculations, the nuclear strength is ultimately fixed by the Bohr-Sommerfeld quantization condition \cite{Buck1992,Kelkar2007} for the quasibound $\alpha$-cluster state,
\begin{equation}\label{eq:bs}
\int_{r_1}^{r_2} \sqrt{\frac{2\mu}{\hbar^2}}
\Bigg[
Q_\alpha+\frac{V_0}{1+\exp[(r-R_0)/{a}]} 
-V_C(r)-\frac{C_\ell}{r^2}
\Bigg]^{1/2}dr
=\left(G-\ell+1\right)\frac{\pi}{2},
\end{equation}
where $Q_\alpha$ is the decay energy, $r_1$ and $r_2$ are the first and second turning points, and $G$ is the global quantum number originated from the Pauli exclusion principle. The value of $G$ follows from the Wildermuth-Tang rule,
\begin{equation}\label{eq:wildermuth}
G=2N_r+\ell=\sum_{i=1}^{4}(2n_i+\ell_i)=\sum_{i=1}^{4}g_i,
\end{equation}
where $N_r$ is the number of internal radial nodes and $g_i=2n_i+\ell_i$ are the oscillator quanta assigned to the four nucleons forming the $\alpha$ cluster \cite{Wildermuth1977,Seif2020Wildermuth,CHIEN2022,Buck1992}. For the present favored $\ell=0$ decays, Eq.~(\ref{eq:wildermuth}) reduces to $N_r=G/2$. We adopt the standard $\alpha$-decay shell prescription in which the cluster nucleons occupy Pauli-allowed oscillator shells above those filled in the daughter nucleus, with $g_i=4$, 5, and 6 for the 50--82, 82--126, and above-126 shells, respectively \cite{Qian2014}. 

\subsection{Two-potential approach and modified TPA}\label{subsec:tpa}

\begin{figure}[t]
	\centering
	\includegraphics[width=0.6\linewidth]{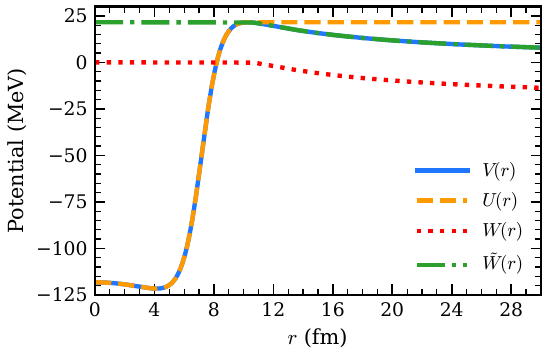}
	\caption{Potential splitting in the two-potential approach for the $^{208}$Pb$+\alpha$ system. The physical potential $V(r)$ is separated into the inner potential $U(r)$ and the perturbation $W(r)$ at the separation radius $R_s$. The shifted potential $\widetilde{W}(r)=W(r)+V_s$ enters the scattering equation for $\chi_{k_0}(r)$.}
	\label{fig:tpa_potential}
\end{figure}

The TPA divides the physical interaction into an inner potential $U(r)$ and an outer perturbation $W(r)$ at a separation radius $R_s$ \cite{Gurvitz1987,Gurvitz1988}. In practice, $R_s$ is the point at which the final physical potential reaches its barrier maximum after the nuclear potential depth has been determined. The split reads
\begin{equation}
V(r)=U(r)+W(r),
\end{equation}
with
\begin{equation}
U(r)=
\begin{cases}
V(r), & r\le R_s,\\
V_s, & r>R_s,
\end{cases}
\end{equation}
\begin{equation}
W(r)=
\begin{cases}
0, & r\le R_s,\\
V(r)-V_s, & r>R_s,
\end{cases}
\end{equation}
where $V_s=V(R_s)$. All TPA and MTPA expressions below employ the physical centrifugal term of Eq.~(\ref{eq:vtotal}). Fig.~\ref{fig:tpa_potential} illustrates the potential splitting for the representative $^{208}$Pb$+\alpha$ system.

The bound-state wave function $u_b(r)$ satisfies the inner eigenvalue equation,
\begin{equation}
\left[-\frac{\hbar^2}{2\mu}\frac{d^2}{dr^2}+U(r)\right]u_b(r)=E_0u_b(r),
\end{equation}
with $u_b(0)=0$ and $\int_0^\infty |u_b(r)|^2dr=1$. The energy is measured from the $\alpha$-daughter threshold and set to $E_0=Q_\alpha$ in the numerical work. The regular scattering solution $\chi_{k_0}(r)$ satisfies
\begin{equation}
\left[-\frac{\hbar^2}{2\mu}\frac{d^2}{dr^2}+\widetilde{W}(r)\right]\chi_{k_0}(r)=E_0\chi_{k_0}(r),
\end{equation}
where $\widetilde{W}(r)=W(r)+V_s$, $k_0=\sqrt{2\mu E_0 /\hbar^2}$, and $\eta_C=\mu Z/(\hbar^2 k_0)$. The continuum convention is fixed by the asymptotic standing-wave normalization
\begin{align}\label{eq:chi_norm}
\chi_{k_0}(r\to\infty)
=\cos\delta_\ell\,F_\ell(\eta_C,k_0r)+\sin\delta_\ell\,G_\ell(\eta_C,k_0r),
\end{align}
where $F_\ell$ and $G_\ell$ are the regular and irregular Coulomb functions and $\delta_\ell$ is the nuclear phase shift produced by $\widetilde{W}(r)$. 

With these conventions, the four TPA-type expressions for the width, corresponding to the volume, the two surface, and boundary representations, are \cite{Gurvitz1987,Gurvitz1988,Gurvitz2004}
\begin{align}
\Gamma_{\rm TPA1}
&=\frac{4\mu}{\hbar^2k_0}
\left|\int_{R_s}^{\infty}u_b(r)W(r)\chi_{k_0}(r)\,dr\right|^2, \label{eq:tpa1}\\
\Gamma_{\rm TPA2}
&=\frac{\hbar^2}{\mu k_0}
\left|u_b(R_s)\left[\beta_s\chi_{k_0}(R_s)+\chi_{k_0}'(R_s)\right]\right|^2, \label{eq:tpa2}\\
\Gamma_{\rm TPA3}
&=\frac{\hbar^2}{\mu k_0}
\left|u_b(R_s)\chi_{k_0}'(R_s)-\chi_{k_0}(R_s)u_b'(R_s)\right|^2, \label{eq:tpa3}\\
\Gamma_{\rm TPA4}
&=\frac{4\hbar^2\beta_s^2}{\mu k_0}\left|u_b(R_s)\chi_{k_0}(R_s)\right|^2, \label{eq:tpa4}
\end{align}
with $\beta_s={\sqrt{2\mu (V_s-E_0)}}/{\hbar}$.
Equations~(\ref{eq:tpa1}), (\ref{eq:tpa2}), and (\ref{eq:tpa3}) are algebraically equivalent within the Gurvitz construction, whereas Eq.~(\ref{eq:tpa4}) is the simplified boundary expression obtained when $\chi_{k_0}'(R_s)\approx \beta_s \chi_{k_0}(R_s)$ under the barrier \cite{Gurvitz1987,Gurvitz1988}. It should be noted that all four formulas use the normalized inner bound-state wave function $u_b$, not the outgoing Gamow (resonant) wave function employed in the DW and direct methods \cite{Aberg1997,Davids2000,Alavi2024}. This is an advantage of the TPA methods over more exact ones since the normalized wave function $u_b$ is considerably more straightforward to obtain than its resonant counterpart. 

\begin{figure}[t]
	\centering
	\includegraphics[width=\linewidth]{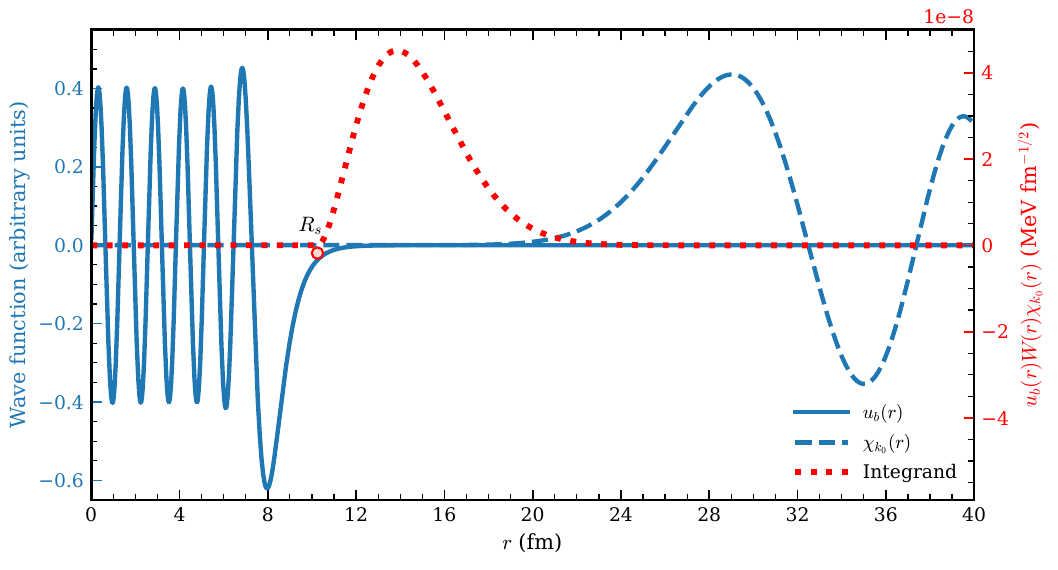}
	\caption{Representative overlap structure entering the volume-integral TPA width for the $^{208}$Pb$+\alpha$ system. The left vertical scale shows the wave functions $u_b(r)$ and $\chi_{k_0}(r)$, with $\chi_{k_0}$ divided by 4 for display. The right vertical scale shows the kernel $u_b(r)W(r)\chi_{k_0}(r)$.}
	\label{fig:overlap}
\end{figure}

Fig.~\ref{fig:overlap} displays the representative overlap structure for the $^{208}$Pb$+\alpha$ system. The bound-state wave function $u_b(r)$ falls off rapidly through the barrier, whereas the regular scattering wave function $\chi_{k_0}(r)$ grows in importance outside the inner region. Their product, weighted by $W(r)$, concentrates near the barrier surface. This localization explains why the volume integral in Eq.~(\ref{eq:tpa1}) can be faithfully reproduced by the surface (\ref{eq:tpa2})--(\ref{eq:tpa3}) and boundary expressions (\ref{eq:tpa4}).

Our comparison also includes the modified two-potential approach (MTPA) of Gurvitz \textit{et al.} \cite{Gurvitz2004}, which replaces the sharp TPA split at $R_s$ with a matching construction in the barrier tail and therefore provides the nontrivial test within the TPA family. The matching radius $\bar r$ is chosen where the total potential tail is well approximated by the repulsive part
\begin{equation}
V_{\rm rep}(r)=V_C(r)+\frac{\hbar^2\ell(\ell+1)}{2\mu r^2}.
\end{equation}
The matching window extends from the barrier-top radius $R_b$ to a separation radius $R_{\rm sep}$ located between the barrier top and the outer turning point $r_3$, where $V(r_3)=Q_\alpha$. We choose $R_{\rm sep}$ such that the inner eigenenergy agrees with $Q_\alpha$ to within $10^{-3}$ MeV. Within this window, $\bar r$ is placed where the attractive nuclear tail is small relative to the Coulomb-plus-centrifugal reference as
\begin{equation}
\left|1-\frac{V_{\rm rep}(r)}{V(r)}\right| \leq 10^{-5}.
\end{equation}
The MTPA matching condition then reads
\begin{equation}\label{eq:mtpa_match}
\frac{u_b'(\bar r)}{u_b(\bar r)}
=\frac{d}{dr}\ln G_{\ell}(\eta_C,k_0r)\bigg|_{r=\bar r},
\end{equation}
In the present benchmark, the energy is fixed at $E_0=Q_\alpha$, the inner-potential strength is adjusted until the inner eigenvalue matches $Q_\alpha$, $\bar r$ is selected by the tail criterion above, and the width is evaluated from the simplified irregular-Coulomb expression
\begin{equation}\label{eq:mtpa}
\Gamma_{\rm MTPA}=\frac{\hbar^2k_0}{\mu}
\left|\frac{u_b(\bar r)}{G_{\ell}(\eta_C,k_0\bar r)}\right|^2.
\end{equation}
The matching radius is thus constrained by the barrier geometry rather than to an arbitrary value \cite{Gurvitz2004}. Equation~(\ref{eq:mtpa}) is the simplified MTPA width obtained after neglecting the exponentially small regular-Coulomb and phase-shift corrections in the barrier region. As shown below, the small MTPA envelope confirms that the selected tail matching points remain on the same accuracy order as the more exact TPA forms.

\subsection{WKB penetrability and prefactors}\label{subsec:wkb}

For the WKB comparison we retain the QCPM input prescription but evaluate the nuclear potential depth and the turning points with the Langer centrifugal term. The effective potential becomes
\begin{equation}\label{eq:wkb_potential}
V_{\rm WKB}(r)=V_N(r)+V_C(r)+
\frac{C_\ell}{r^2}.
\end{equation}
The three turning points $r_1<r_2<r_3$ satisfy $V_{\rm WKB}(r_i)=Q_\alpha$. The interval $r_1<r<r_2$ is the inner classically allowed region and $r_2<r<r_3$ is the classically forbidden region inside the barrier. The use of Langer modification in Eq.~(\ref{eq:clanger}) is the standard radial WKB prescription to ensure correct radial boundary behavior. This modified centrifugal term is distinct from the physical centrifugal term used in the TPA/MTPA wave functions. Consequently, the effective potential in the WKB construction need not coincide with the corresponding TPA/MTPA one. 

The WKB width takes the form
\begin{equation}\label{eq:wkb_master}
\Gamma_{\rm WKB}^{(j)}=\hbar\nu_j P_{\rm WKB},
\qquad
j={\rm GK,RT,HO,CF},
\end{equation}
where all four frequency (or prefactor) prescriptions share the same penetration probability,
\begin{align}
P_{\rm WKB}
&=\left\{1+\exp\left[2\int_{r_2}^{r_3}\kappa(r)\,dr\right]\right\}^{-1},
\label{eq:wkb_penetrability}\\
\kappa(r)
&=\sqrt{\frac{2\mu}{\hbar^2}\left[V_{\rm WKB}(r)-Q_\alpha\right]} . \nonumber
\end{align}
The comparison between the four WKB prescriptions in this work thus reflects the sole effect of the assault frequency or normalization prefactor.

The first choice is the Gurvitz-Kalbermann (GK) normalization of the WKB wave function with the TPA quasibound state in the inner pocket \cite{Gurvitz1987,Gurvitz1988}
\begin{equation}\label{eq:nu_gk}
\nu_{\rm GK}=\frac{\hbar\mathcal{N}_{\rm GK}}{4\mu},
\qquad
\Gamma_{\rm WKB}^{({\rm GK})}
=\frac{\hbar^2\mathcal{N}_{\rm GK}}{4\mu}P_{\rm WKB},
\end{equation}
with
\begin{equation}\label{eq:norm_gk}
\mathcal{N}^{-1}_{\rm GK}
=\int_{r_1}^{r_2}
\frac{dr}{k_{\rm in}(r)}
\cos^2\left[
\int_{r_1}^{r}k_{\rm in}(r')\,dr'-\frac{\pi}{4}
\right],
\end{equation}
where
\begin{equation}
k_{\rm in}(r)
=\sqrt{\frac{2\mu}{\hbar^2}\left[Q_\alpha-V_{\rm WKB}(r)\right]}.
\end{equation}

Replacing the squared cosine term in Eq.~(\ref{eq:norm_gk}) by its averaged value $1/2$ reduces Eq.~(\ref{eq:nu_gk}) to the familiar round-trip (RT) frequency \cite{Gurvitz1987,Gurvitz1988}
\begin{align}\label{eq:nu_rt}
\nu_{\rm RT}
&=\frac{\hbar}{2\mu}
\left[\int_{r_1}^{r_2}\frac{dr}{k_{\rm in}(r)}\right]^{-1}.
\end{align}
This form follows from the classical period of motion in the inner allowed pocket and is the widely used prefactor associated with the WKB quantization picture \cite{Kelkar2007,CHIEN2022,Buck1992}. Although Eq.~(\ref{eq:nu_gk}) appears to retain more of the TPA wave structure than Eq.~(\ref{eq:nu_rt}), the two prescriptions are better regarded as competing approximations to the normalization of the same exact quasibound state. As shown in Section~\ref{sec:results}, the averaged form turns out to provide the more faithful representation of the two.

Another popular WKB frequency is the harmonic-oscillator (HO) prescription
\begin{equation}\label{eq:nu_ho}
\nu_{\rm HO}
=\frac{(G+3/2)\hbar}{1.2\pi \mu R_0^2},
\end{equation}
where $G$ is the global quantum number and $R_0$ is the WS radius parameter of the present QCPM potential \cite{CHIEN2022,Sharma2022frequency,Dong2010,Zhang2011}.

Finally, the classical-frequency (CF) frequency models the emitted $\alpha$ particle as bouncing back and forth across the nuclear well with kinetic energy $Q_\alpha$,
\begin{equation}\label{eq:nu_cf}
\nu_{\rm CF}
=\frac{1}{R_0}\sqrt{\frac{Q_\alpha}{2\mu}} .
\end{equation}
These classical-frequency prescriptions appear frequently in generalized liquid drop model (GLDM) and preformed-cluster calculations \cite{CHIEN2022,Sharma2022frequency,Bao2014GLDM,Deng2021GLDM,Zhang2011}.

\section{Results and discussion}\label{sec:results}

\subsection{Comparison between models}
The numerical comparison in this work covers 161 even-even parent nuclei with $A_p=144$--252, all treated as favored $\ell=0$ ground-state transitions with the QCPM input of Sec.~\ref{subsec:potential}. Decay energies are taken from the AME2020 evaluation \cite{Wang2021} and spin-parity assignments from NUBASE2020 \cite{Kondev2021}.

\begin{figure}[t]
\centering
\includegraphics[width=0.92\textwidth]{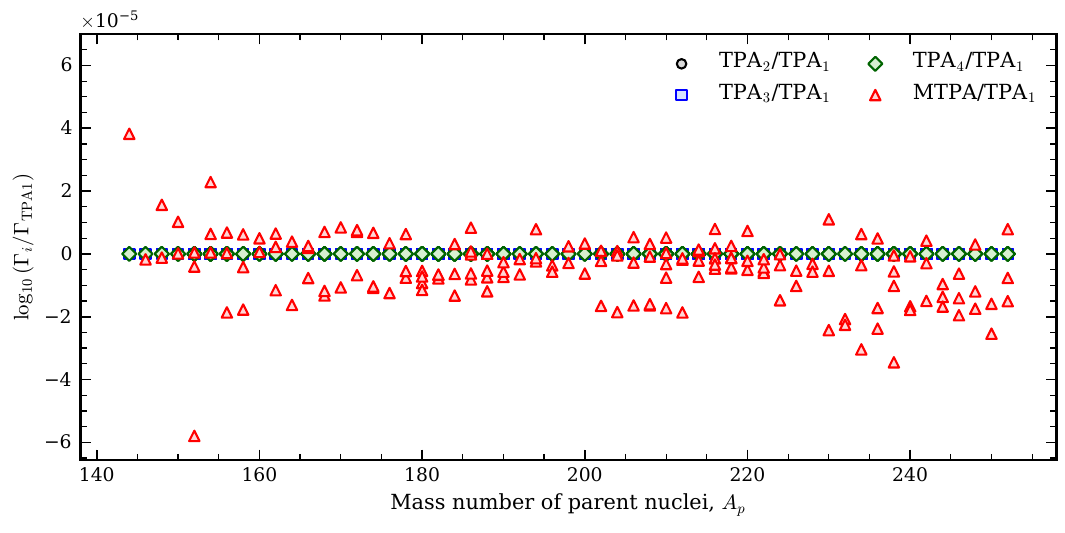}
\caption{Logarithmic deviations among the TPA-type widths, shown as $\log_{10}(\Gamma_i/\Gamma_{\rm TPA1})$ with $i={\rm TPA2,TPA3,TPA4,MTPA}$ for 161 even-even parent nuclei.}
\label{fig:logTPA}
\end{figure}

We begin with the internal consistency of the TPA family. Fig.~\ref{fig:logTPA} presents the width-level comparison within the TPA family in terms of $\log_{10}(\Gamma_i/\Gamma_{\rm TPA1})$, where the volume-integral result $\Gamma_{\rm TPA1}$ serves as the reference. For the 161 nuclei considered, the four TPA formulas are visually indistinguishable on the scale of the figure. The TPA2 deviations range from $-8.02\times 10^{-9}$ to $1.18\times 10^{-8}$, those of TPA3 from $-3.28\times 10^{-9}$ to $1.01\times 10^{-8}$, and those of TPA4 from $-6.19\times 10^{-9}$ to $7.28\times 10^{-9}$. The MTPA results exhibit the largest spread in the comparison, ranging from $-5.79\times 10^{-5}$ to $3.82\times 10^{-5}$.

These minuscule differences confirm that the TPA-type calculations define a stable width reference for the adopted input. The volume integral~(\ref{eq:tpa1}), the algebraically equivalent surface forms~(\ref{eq:tpa2})--(\ref{eq:tpa3}), and the reduced boundary approximation~(\ref{eq:tpa4}) all agree to better than few parts in $10^{-8}$ of logarithmic difference. The MTPA matching form~(\ref{eq:mtpa}) provides the nontrivial $\alpha$-decay extension, since it probes whether a barrier-region matching construction remains on the same quantum width scale as the sharp potential split of the original TPA. The observed agreement extends the method-comparison findings of \AA{}berg \textit{et al.} for spherical proton emitters and of Alavi \textit{et al.} for uranium $\alpha$ decay \cite{Aberg1997,Alavi2024}, according to which different quantum width formulas coincide in spherical one-body calculations once the potential, normalization, and matching conventions are controlled. That the same convergence is obtained here with a QCPM interaction, rather than the double-folding potential of Ref.~\cite{Alavi2024}, indicates that it is intrinsic to the TPA construction and not an effect of a particular nuclear interaction.

\begin{figure}[t]
\centering
\includegraphics[width=0.92\textwidth]{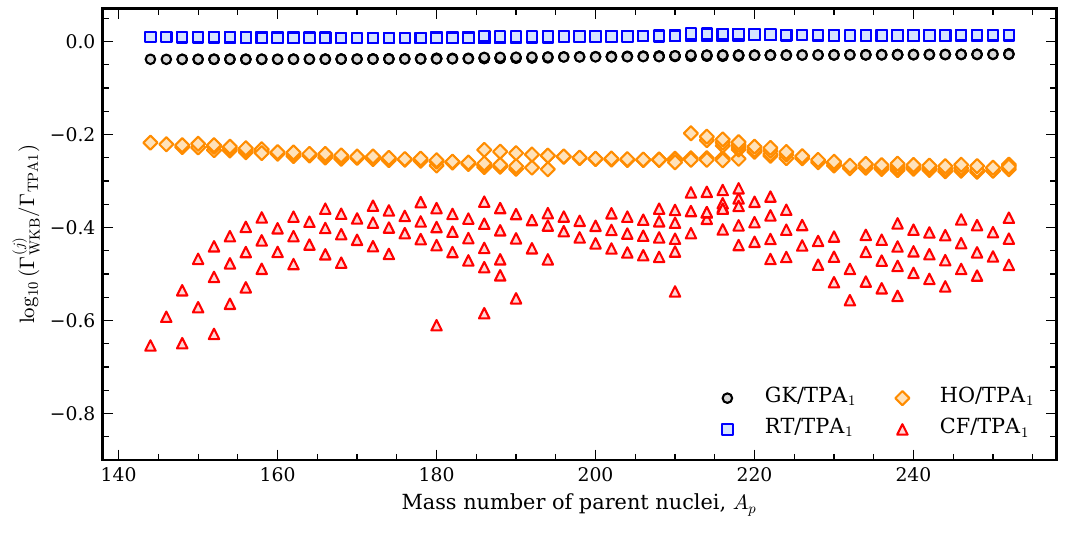}
\caption{Comparison of four WKB widths with the volume-integral TPA width, shown as $\log_{10}(\Gamma_{\rm WKB}^{(j)}/\Gamma_{\rm TPA1})$ for $j={\rm GK,RT,HO,CF}$ in 161 even-even parent nuclei.}
\label{fig:TPA_WKB_scatter}
\end{figure}

We now confront this quantum reference with the semiclassical WKB widths. Fig.~\ref{fig:TPA_WKB_scatter} shows $\log_{10}(\Gamma_{\rm WKB}^{(j)}/\Gamma_{\rm TPA1})$ as a function of the parent mass number for the four prefactor prescriptions of Sec.~\ref{subsec:wkb}. Since the four prescriptions share the same penetrability~(\ref{eq:wkb_penetrability}) under the adopted Langer convention, the systematic displacements between the bands in Fig.~\ref{fig:TPA_WKB_scatter} directly measure the prefactor dependence of the absolute width scale.

The central positive result is that the round-trip (RT) frequency tracks the TPA1 width scale closely. It gives $\log_{10}(\Gamma_{\rm WKB}^{({\rm RT})}/\Gamma_{\rm TPA1})$ between $7.71\times 10^{-3}$ and $1.74\times 10^{-2}$ in the 161-nucleus survey. This small positive offset is partly attributable to the Langer centrifugal coefficient entering the WKB turning-point integrals, as opposed to the physical centrifugal term retained in the TPA wave functions, compounded by high-order barrier-top corrections to the action integral. The GK prefactor yields slightly smaller widths, with logarithmic deviations from $-3.84\times 10^{-2}$ to $-2.62\times 10^{-2}$. The HO and CF estimates, in contrast, exhibit pronounced negative shifts, with logarithmic deviations between $-2.80\times 10^{-1}$ and $-1.97\times 10^{-1}$ for the HO prescription and between $-6.54\times 10^{-1}$ and $-3.16\times 10^{-1}$ for the CF prescription.

This deviation ordering is physically transparent. The action integral fixes the dominant exponential dependence of the width, so the inner normalization, or the related assault frequency, shifts the absolute single-particle width by precisely the logarithmic offsets shown above. The relative behavior of $\nu_{\rm RT}$ and $\nu_{\rm GK}$ with respect to TPA1 follows the same pattern reported in the erratum of \AA{}berg \textit{et al.} for proton decay, where the averaged-normalization (RT) WKB result agrees more closely with the quantum reference than the GK normalization \cite{Aberg1998Erratum,Aberg1997}. The present work establishes the analogous $\alpha$-decay statement and carries the comparison one step further. Once the equivalence among quantum methods (TPA, DW, direct) is secured, as demonstrated by Alavi \textit{et al.}, the remaining practical question is which semiclassical prefactor in WKB recovers that quantum width scale over a broad dataset. The WKB approximation provides a reliable tool with excellent agreement with TPA formalism when the RT prefactor is used, but this prefactor choice is by no means arbitrary. A WKB width is meaningful as a benchmark quantity only when the prefactor and the radial quantization convention are specified together with the penetrability \cite{Aberg1997,Aberg1998Erratum,Kelkar2007,CHIEN2022,Sharma2022frequency}.


To probe a wider region of the nuclear chart in more detail, Table~\ref{tab:benchmark} presents a separate illustrative 10-nucleus run spanning $A_p=104$--294, covering from light superallowed up to superheavy $\alpha$ emitters. This set is not a subset of the 161-nucleus comparison, which is restricted to $A_p=144$--252, but the same input convention and calculation routine applies. We emphasize that all tabulated values are single-particle decay widths with no preformation factor applied. The scope of the benchmark is deliberately restricted. Static deformation, channel coupling, odd-particle structure, and phenomenological $\alpha$ preformation are excluded. These constraints keep the comparison focused on method dependence at the single-particle-width level. 

\begin{table}[!t]
	\caption{Single-particle decay widths $\Gamma$ in MeV for the illustrative 10 nuclei. The TPA and MTPA columns are taken from the two-potential approach calculations, while the WKB columns use the same penetrability with the indicated prefactor.}
	\label{tab:benchmark}
	\centering
	\renewcommand{\arraystretch}{1.15}
	\setlength{\tabcolsep}{3.0pt}
	\resizebox{\textwidth}{!}{%
		\begin{tabular}{lccccccccc}
			\toprule
			\multirow{2}{*}{Nuclei} & \multicolumn{5}{c}{TPA width} & \multicolumn{4}{c}{WKB width} \\
			\cmidrule(lr){2-6}\cmidrule(lr){7-10}
			& TPA1 & TPA2 & TPA3 & TPA4 & MTPA & GK & RT & HO & CF \\
			\midrule
			$^{104}$Te & $3.672\times 10^{-14}$ & $3.672\times 10^{-14}$ & $3.672\times 10^{-14}$ & $3.672\times 10^{-14}$ & $3.672\times 10^{-14}$ & $3.323\times 10^{-14}$ & $3.758\times 10^{-14}$ & $2.534\times 10^{-14}$ & $1.549\times 10^{-14}$ \\
			$^{108}$Xe & $3.387\times 10^{-17}$ & $3.387\times 10^{-17}$ & $3.387\times 10^{-17}$ & $3.387\times 10^{-17}$ & $3.387\times 10^{-17}$ & $3.058\times 10^{-17}$ & $3.445\times 10^{-17}$ & $2.205\times 10^{-17}$ & $1.300\times 10^{-17}$ \\
			$^{144}$Nd & $8.952\times 10^{-45}$ & $8.952\times 10^{-45}$ & $8.952\times 10^{-45}$ & $8.952\times 10^{-45}$ & $8.952\times 10^{-45}$ & $8.198\times 10^{-45}$ & $9.155\times 10^{-45}$ & $5.425\times 10^{-45}$ & $1.987\times 10^{-45}$ \\
			$^{180}$W & $5.033\times 10^{-47}$ & $5.033\times 10^{-47}$ & $5.033\times 10^{-47}$ & $5.033\times 10^{-47}$ & $5.033\times 10^{-47}$ & $4.633\times 10^{-47}$ & $5.130\times 10^{-47}$ & $2.724\times 10^{-47}$ & $1.235\times 10^{-47}$ \\
			$^{212}$Po & $1.745\times 10^{-14}$ & $1.745\times 10^{-14}$ & $1.745\times 10^{-14}$ & $1.745\times 10^{-14}$ & $1.745\times 10^{-14}$ & $1.632\times 10^{-14}$ & $1.817\times 10^{-14}$ & $1.107\times 10^{-14}$ & $8.263\times 10^{-15}$ \\
			$^{226}$Ra & $1.214\times 10^{-32}$ & $1.214\times 10^{-32}$ & $1.214\times 10^{-32}$ & $1.214\times 10^{-32}$ & $1.214\times 10^{-32}$ & $1.135\times 10^{-32}$ & $1.250\times 10^{-32}$ & $6.677\times 10^{-33}$ & $3.764\times 10^{-33}$ \\
			$^{238}$U & $1.868\times 10^{-39}$ & $1.868\times 10^{-39}$ & $1.868\times 10^{-39}$ & $1.868\times 10^{-39}$ & $1.868\times 10^{-39}$ & $1.751\times 10^{-39}$ & $1.922\times 10^{-39}$ & $9.883\times 10^{-40}$ & $5.304\times 10^{-40}$ \\
			$^{270}$Hs & $2.664\times 10^{-22}$ & $2.664\times 10^{-22}$ & $2.664\times 10^{-22}$ & $2.664\times 10^{-22}$ & $2.664\times 10^{-22}$ & $2.515\times 10^{-22}$ & $2.758\times 10^{-22}$ & $1.437\times 10^{-22}$ & $1.083\times 10^{-22}$ \\
			$^{286}$Fl & $1.752\times 10^{-20}$ & $1.752\times 10^{-20}$ & $1.752\times 10^{-20}$ & $1.752\times 10^{-20}$ & $1.752\times 10^{-20}$ & $1.656\times 10^{-20}$ & $1.813\times 10^{-20}$ & $9.322\times 10^{-21}$ & $7.658\times 10^{-21}$ \\
			$^{294}$Og & $6.309\times 10^{-18}$ & $6.309\times 10^{-18}$ & $6.309\times 10^{-18}$ & $6.309\times 10^{-18}$ & $6.309\times 10^{-18}$ & $5.969\times 10^{-18}$ & $6.541\times 10^{-18}$ & $3.380\times 10^{-18}$ & $3.000\times 10^{-18}$ \\
			\bottomrule
		\end{tabular}
	}
\end{table}

The ten TPA-family widths span nearly 33 orders of magnitude, from the extremely hindered decays of $^{180}$W and $^{144}$Nd to the much shorter-lived emitters $^{104}$Te and $^{212}$Po and the superheavy nuclei. Across this entire range, the five TPA-type columns remain identical to the displayed precision, which confirms that the close agreement in Fig.~\ref{fig:logTPA} is not an artifact of one narrow mass interval or one particular decay energy.

The WKB columns display the systematic hierarchy anticipated by Fig.~\ref{fig:TPA_WKB_scatter}. Relative to TPA1, the RT widths are higher by about $1.73$--$4.09\%$, whereas the GK widths are lower by about $5.38$--$9.72\%$, the HO widths by about $30.99$--$47.11\%$, and the CF widths by about $52.44$--$77.80\%$. These differences reflect how each prescription represents the motion in the inner pocket before barrier penetration. Because all four WKB columns share the same penetrability, the table isolates the prefactor dependence in a way that a half-life comparison involving fitted preformation factors would normally obscure.

This benchmark table therefore serves both as a reproducibility check and as a scale diagnostic. It provides concrete values for comparing independent evaluations of the TPA formulas, MTPA matching, and WKB prefactors, and it demonstrates how much the frequency prescription impacts the WKB width. WKB prescriptions sharing the same penetration probability acquire different absolute scales when the prefactor is changed. The contrast between the stable TPA-family and the larger WKB-prefactor shifts is especially relevant for superheavy and neutron-deficient nuclei, where a disagreement with experiment may reflect preformation physics, deformation, or spectroscopy rather than a deficiency in the tunneling calculation itself.

\subsection{Uniform Airy analysis of the WKB prefactors}

It remains to explain why the averaged (RT) prescription outperforms the seemingly more complete GK one as first reported in Ref.~\cite{Aberg1997,Aberg1998Erratum} and confirmed in our benchmark. The consistent ordering $\Gamma^{\rm (GK)}_{\rm WKB} < \Gamma^{\rm (RT)}_{\rm WKB} \simeq \Gamma_{\rm TPA1}$
can be traced to the normalization of the quasibound state in the inner pocket. In the TPA, the width is built from the exactly normalized bound state $u_b$, and the WKB prefactor must reproduce the barrier amplitude of this exact state. Away from the turning points, the leading WKB form in the inner allowed region is
\begin{equation}
	u_b(r)\simeq u_{\rm WKB}(r)= \sqrt{\frac{\mathcal N}{k_{\rm in}(r)}} \cos\left[S(r)-\frac{\pi}{4}\right], \qquad S(r)=\int_{r_1}^{r}k_{\rm in}(r')\,dr',
\end{equation}
so the normalization factor ${\mathcal N}$ of WKB wave function can be determined from the normalization condition of a bound state inside the pocket
\begin{equation}
	\int_{r_1}^{r_2}  \left|u_{\rm WKB}(r)\right|^2 dr
	= {\mathcal N} \int_{r_1}^{r_2} \frac{dr}{k_{\rm in}(r)}
	\cos^2\left[S(r)-\frac{\pi}{4}\right]=1.
\end{equation}
Thus the issue is not whether one should keep or average the oscillations of an already exact WKB wave function. The issue is which prescription gives the correct leading normalization of the exact TPA state. This is nontrivial as the WKB wave function is no longer valid in the region where it crosses the turning point \cite{Berry1972}. 

Using
\begin{equation}
	\cos^2\left(S-\frac{\pi}{4}\right) = \frac12+\frac12\sin 2S,
\end{equation}
for the GK prescription normalizes the literal oscillatory WKB density in the allowed pocket of Eq.~(\ref{eq:norm_gk}) leads to
\begin{equation}\label{eq:ndecomp}
	\mathcal{N}_{\rm GK}^{-1} = \frac12\int_{r_1}^{r_2}\frac{dr}{k_{\rm in}(r)} + \frac12\int_{r_1}^{r_2}\frac{\sin 2S(r)}{k_{\rm in}(r)}\,dr \equiv \mathcal{N}_{\rm RT}^{-1}+\Delta I .
\end{equation}
The first term is the round-trip (RT),averaged, or classical-period normalization in Eq.~(\ref{eq:nu_rt})
\begin{equation}
	\mathcal N_{\rm RT}^{-1} = \frac12\int_{r_1}^{r_2}\frac{dr}{k_{\rm in}(r)} =\dfrac{\hbar}{4\mu \nu_{\rm RT}}.
\end{equation}
The second term,
\begin{equation}
	\Delta I = \frac12\int_{r_1}^{r_2}\frac{\sin 2S(r)}{k_{\rm in}(r)}\,dr ,
\end{equation}
quantifies the difference between the GK and RT normalizations and is \emph{not} a controlled higher-order correction to the exact TPA norm. In the interior of the pocket it averages away, but near a turning point ($k_{\rm in}\to 0$), the factor ($1/k_{\rm in}$) enhances the endpoint contribution. This is precisely the region where the ordinary WKB amplitude ($k_{\rm in}^{-1/2}$) is nonuniform and must be replaced by an Airy approximation \cite{Berry1972}.

To quantify the difference between the RT and GK normalizations, we isolate one ordinary linear turning point $r_t$. Let $x$ be the distance measured from the turning point into the allowed region ($x=0$ is the turning point). Then, locally,
\begin{equation}
	E-V_{\rm WKB}(r)\simeq Fx, \qquad F=|V'_{\rm WKB}(r_t)|,
\end{equation}
and
\begin{equation}
	k(x)=\frac{\sqrt{2\mu Fx}}{\hbar} = \frac{\sqrt{x}}{\ell_{\rm tp}^{3/2}}, \qquad \ell_{\rm tp} = \left(\frac{\hbar^2}{2\mu F}\right)^{1/3}.
\end{equation}
The local action from the turning point is
\begin{equation}
	S(x) = \int_0^x k(x')\,dx' = \frac23\left(\frac{x}{\ell_{\rm tp}}\right)^{3/2}.
\end{equation}
The endpoint part of the oscillatory GK correction is therefore, for the standard turning-point phase,
\begin{equation}\label{eq:deltaI}
	\Delta I_{\rm tp} = \frac12\int_0^\infty \frac{dx}{k(x)}\sin[2S(x)] = \frac{\Gamma(1/3)}{6} \left(\frac34\right)^{1/3} \ell_{\rm tp}^2 \approx 0.406\,\ell_{\rm tp}^2 .
\end{equation}
Thus $\Delta I_{\rm tp}$ is the extra endpoint term obtained by subtracting the smooth averaged density from the literal WKB density. In the present $\alpha$-decay calculation the inner Langer turning point is very stiff, so its Airy length is small while the dominant soft contribution comes from the outer pocket turning point. This gives a positive $\Delta I$, hence
\begin{equation}
	\mathcal N_{\rm GK}^{-1} > \mathcal N_{\rm RT}^{-1}, \qquad \Gamma^{\rm (GK)}_{\rm WKB} < \Gamma^{\rm (RT)}_{\rm WKB}.
\end{equation}

The crucial point is that the exact TPA wave function does not contain this GK endpoint term. Near the same linear turning point, the exact Schr\"odinger equation reduces to the Airy equation. Introducing the Airy coordinate
\begin{equation}
	t=\frac{r-r_t}{\ell_{\rm tp}},
\end{equation}
so that $t<0$ is the allowed side and $t>0$ is the forbidden side. The uniform local solution, which is the proper form of the TPA wave function near the turning point, is presented in terms of the Airy function
\begin{equation}
	u_{\rm TPA}(r) = u_{\rm Ai}(r)=C\,{\rm Ai}(t).
\end{equation}
On the allowed side, write $t=-X$, where $X=x/\ell_{\rm tp}$. For $X\gg1$,
\begin{equation}
	{\rm Ai}(-X) \simeq \frac{1}{\sqrt{\pi}}X^{-1/4} \sin\left(\frac{2}{3}X^{3/2}+\frac{\pi}{4}\right).
\end{equation}
Since $k(x)={\sqrt{x}}/{\ell_{\rm tp}^{3/2}}$ and ${1}/{\sqrt{k(x)}} = \ell_{\rm tp}^{1/2}X^{-1/4}$,
matching the Airy solution to the WKB envelope gives
\begin{equation}\label{eq:airy_wkb_match}
	C\,{\rm Ai}(-X) \simeq \sqrt{\frac{\mathcal N}{k(x)}} \sin\left(S(x)+\frac{\pi}{4}\right), \qquad C^2=\pi\ell_{\rm tp}\mathcal N .
\end{equation}
Now compute the exact local norm of this Airy wave over an interval extending a distance $x=\ell_{\rm tp}X$ into the allowed region and including the forbidden tail
\begin{equation}
	\int_{-X}^{\infty}|u_{\rm Ai}|^2\,dr = C^2\ell_{\rm tp} \int_{-X}^{\infty}{\rm Ai}^2(t)\,dt .
\end{equation}
The Airy identity
\begin{equation}
	\frac{d}{dt} \left[ t\,{\rm Ai}^2(t)-{\rm Ai}'^2(t) \right] = {\rm Ai}^2(t),
\end{equation}
gives
\begin{equation}\label{eq:airy_norm}
	\int_{-X}^{\infty}{\rm Ai}^2(t)\,dt = X\,{\rm Ai}^2(-X)+{\rm Ai}'^2(-X) = \frac{\sqrt{X}}{\pi} +\mathcal O(X^{-1}) .
\end{equation}
Using $C^2=\pi\ell_{\rm tp}\mathcal N$, Eq.~\eqref{eq:airy_norm} becomes
\begin{equation}\label{eq:airy_to_rt}
	\int_{-X}^{\infty}|u_{\rm Ai}|^2\,dr = \mathcal N\,\ell_{\rm tp}^2\sqrt{X} +\mathcal O(\mathcal N\ell_{\rm tp}^2X^{-1}).
\end{equation}
We also have the averaged (RT) norm over the same allowed-side interval is
\begin{equation}
	\mathcal N \frac12\int_0^{\ell_{\rm tp}X}\frac{dx}{k(x)} = \mathcal N \frac12\int_0^{\ell_{\rm tp}X} \frac{dx}{\sqrt{x}/\ell_{\rm tp}^{3/2}} = \mathcal N\,\ell_{\rm tp}^2\sqrt{X}.
\end{equation}
Therefore
\begin{equation}\label{eq:airy_equals_rt}
		\int_{-X}^{\infty}|u_{\rm Ai}|^2\,dr = \mathcal N \frac12\int_0^{\ell_{\rm tp}X}\frac{dx}{k(x)} +\mathcal O(\mathcal N\ell_{\rm tp}^2X^{-1}) .
\end{equation}
This equation shows the direct connection between the proper Airy function and the RT prefactor. The exact uniform turning-point norm equals the cosine-averaged (RT) form of the WKB prefactor at leading order.

The same conclusion can also be seen by separating the Airy norm into the classically allowed and forbidden parts. In dimensionless Airy units,
\begin{equation}
	\int_{-X}^{\infty}{\rm Ai}^2(t)\,dt = 	\int_{-X}^{0}{\rm Ai}^2(t)\,dt+	\int_0^\infty{\rm Ai}^2(t)\,dt,
\end{equation}
with the allowed-side integral is
\begin{equation}
	\int_{-X}^{0}{\rm Ai}^2(t)\,dt = \frac{\sqrt{X}}{\pi} - {\rm Ai}'^2(0) +\mathcal O(X^{-1}),
\end{equation}
whereas the forbidden tail is
\begin{equation}
	\int_0^\infty{\rm Ai}^2(t)\,dt = {\rm Ai}'^2(0).
\end{equation}
The genuine allowed-side turning-point contribution is therefore canceled by the forbidden tail. The RT prescription omits both pieces and hence retains the correct leading term of ${\sqrt{X}}/{\pi}$. The seemingly more complete GK prescription, by contrast, keeps a divergent endpoint term generated by the nonuniform WKB density but does not include the compensating forbidden-tail structure of the exact Airy solution.

Consequently, the leading semiclassical WKB normalization of the exact TPA bound state is
\begin{equation}
	\mathcal N_{\rm TPA}^{-1} = \mathcal N_{\rm RT}^{-1} + \text{higher-order uniform-WKB corrections}.
\end{equation}
Thus the replacement $\cos^2\to 1/2$, which results in the commonly used Eq.~\ref{eq:nu_rt}, is not merely a crude numerical simplification. It is the leading uniform semiclassical normalization of the exact quasibound state entering the TPA width, equivalent to the classical sojourn-time density \cite{Reinisch1997} and to $\mathcal{N}_{\rm RT}=4\mu/(\hbar T)$ with $T$ the round-trip period \cite{Gurvitz1987,Gurvitz1988}.

The correct TPA normalization is not equal to the GK one 
\begin{equation}
	\mathcal N_{\rm GK}^{-1} = \mathcal N_{\rm RT}^{-1} + \Delta I,
\end{equation}
which differs the RT and correct TPA normalization by a spurious contribution $\Delta I$, which itself is a phase-sensitive turning-point artifact of the nonuniform WKB ansatz. Because the spurious term scales as one inverse power of the number of pocket oscillations, the same artifact is much larger for proton emitters with $N_r=0$--$2$, which explains why \AA{}berg \textit{et al.} \cite{Aberg1998Erratum} found the averaged (RT) normalization, rather than the GK normalization, in very close agreement with their DW results. Combined with the known equivalence between DW and TPA methods \cite{Aberg1997,Alavi2024}, our results also explain the good agreement between the DW method and the WKB approximation with RT normalization observed in Ref~.\cite{Bo2026}.

\section{Conclusions}\label{sec:conclusion}

We have benchmarked the single-particle $\alpha$-decay width produced by the TPA and by WKB calculations with four common prefactors, using a fixed QCPM input. For the 161 even-even emitters considered, the four Gurvitz TPA formulas and the MTPA matching expression yield nearly indistinguishable logarithmic widths, so that the TPA/MTPA family provides a stable quantum reference for this potential prescription.

With the adopted Langer convention, the commonly used RT prefactor in the WKB method reproduces the TPA width scale to within $1.7$--$4.1\%$ and is therefore well suited for large $\alpha$-decay surveys. The GK, HO, and CF prescriptions demonstrate that this accuracy depends on the prefactor choice, and WKB widths should be quoted together with the prefactor and the radial quantization convention.

The superior agreement of the RT prefactor over other prescriptions is not trivial. A detailed Airy-function analysis of the quasibound-state norm shows that the $\cos^2\!\to\!1/2$ averaging is the correct leading-order semiclassical normalization of the exact TPA state, whereas the literal GK form carries a spurious positive contribution from the turning-point region where the $1/\sqrt{k_{\rm in}}$ WKB amplitude breaks down. Because this spurious term scales inversely with the number of inner-pocket oscillations, it is much larger for proton emitters with $N_r=0$--2 than for the $\alpha$ states studied here with $N_r\simeq 9$--11. The same mechanism therefore explains the counterintuitive normalization pattern observed by \AA{}berg \textit{et al.} \cite{Aberg1998Erratum,Aberg1997}.

A width-level reference of this kind cleanly separates the tunneling-method scale from the preformation factor contribution, a separation that is especially valuable for heavy and superheavy nuclei identified through their $\alpha$-decay chains. Together with the quantum-method equivalences established by \AA{}berg \textit{et al.} for proton emitters, Alavi \textit{et al.} for uranium $\alpha$ decay, and Bo \cite{Bo2026} for DW-WKB comparison, the present WKB-TPA comparison contribute to a consistent picture of how calculated single-particle decay widths depend on the adopted tunneling formalism across the nuclear chart. The benchmark in this work can be used as a baseline for future extensions to deformation, channel coupling, odd-mass systems, and microscopic preformation factors of $\alpha$-decay theory.

\section*{CRediT authorship contribution statement}

\textbf{Nguyen Thi Huyen Nga}: Conceptualization, Software, Formal analysis, Investigation, Funding acquisition, Writing -- original draft, Writing -- review and editing. \textbf{Le Quoc Thai}: Software, Formal analysis, Investigation, Validation, Visualization, Writing -- review and editing. \textbf{Chau Van Tao}: Resources, Supervision, Validation, Writing -- review and editing. \textbf{Le Hoang Chien}: Conceptualization, Software, Methodology, Validation, Writing -- review and editing. \textbf{Nguyen Tri Toan Phuc}: Conceptualization, Methodology, Software, Formal analysis, Writing -- original draft, Writing -- review and editing.

\section*{Declaration of competing interest}

The authors declare that they have no known competing financial interests or personal relationships that could have appeared to influence the work reported in this paper.

\section*{Data availability}

Data will be made available on request.

\section*{Acknowledgements}

This research is funded by Vietnam National University, Ho Chi Minh City (VNU-HCM) under Grant No. C2026-18-13.

\bibliographystyle{model1a-num-names}
\bibliography{alpha-bibliography}

\end{document}